at the level of a single molecule of mRNA[6]. It is also known that enzymes called RNA helicases can unwind RNA structures adjacent to AUG start codons[7]. However, it was unclear whether cells use uAUG–dsRNA cooperation to dynamically regulate protein synthesis.

To identify functional uAUGs in the plant *Arabidopsis thaliana*, Xiang and colleagues used the approach of sequencing mRNA associated with ribosomes in seedlings. These seedlings were undergoing a defence response triggered by the presence of a fragment, termed elf18, of a bacterial protein called EF-Tu. The authors report that mRNAs that increased their association with ribosomes after treatment with elf18 were enriched in uAUGs that could be used as start sites by ribosomes. Moreover, the authors found that the ribosome density at active uAUGs decreased in response to elf18 compared with the situation during normal growth conditions. These findings suggest that translation preferentially begins at the mAUG through a defence-induced release of uAUG-mediated inhibition.

The authors investigated why some mRNAs might experience a decrease in upstream start codon preference during a defence response mediated by the immune system. Xiang and colleagues identified dsRNA structures near uAUGs, and such structures in mRNAs increased the ribosome occupancy at these start codons under normal growth conditions. Notably, the authors observed consistent dsRNA structures both *in vivo* and *in vitro*, suggesting that they are independent of potential interactions with RNA-binding proteins.

Xiang *et al.* tested whether widespread (global) translational reprogramming was regulated by changes in uAUG-associated structures during the immune response. Indeed, the authors found that dsRNA structures were preferentially unfolded in mRNAs that had increased ribosome occupancy during the immune response. A machine-learning classifier tool helped the authors to identify dsRNA near regulated uAUGs. Cooperation between dsRNA and uAUGs was also observed for the 5′ UTRs of the genes *ATF4* and *BRCA* in human cells, which suggests that translation regulation through uAUG structures is an evolutionarily conserved mechanism found in humans, too.

The authors searched for regulatory factors that unwind dsRNA during the immune response. Xiang and colleagues focused on increased amounts of RNA helicases after treatment with elf18, because these enzymes might act as binding partners, called chaperones, for RNA[8]. The authors found that *A. thaliana* helicases, called RH37-like helicases, which are similar to the yeast protein Ded1p and human protein DDX3X (ref. 9), regulate dsRNA structures near uAUGs during the immune response. Expressing higher than normal levels of RH37-like helicases increased translation from the mAUG for many mRNAs. In plant strains lacking RH37-like helicases, these dsRNA structures persisted during the immune response.

Xiang and colleagues' work has uncovered a previously unknown pathway through which cells can regulate protein synthesis in response to stress conditions by inducing RNA-remodelling enzymes, which remove dsRNA structures to shift from uAUG to mAUG start-codon selection (Fig. 1). This finding was made possible by integrating systematic analyses of translation and of RNA structures.

This research also sheds light on how it is possible to achieve dynamic translation regulation through the arrangement of several regulatory elements in the linear sequence of an mRNA. Furthermore, with mRNA becoming a topic of increasing interest for therapeutic purposes, understanding the regulatory mechanisms that govern translation should provide valuable information for the rational sequence design of mRNA-based treatments.

Given that independent regulatory functions of uAUG, dsRNA and RNA helicases have been reported across different species, the authors propose that the uAUG–dsRNA helicase regulatory module might be widely present in multicellular organisms called eukaryotes. This is partially supported by the authors' experiments examining human cells. However, future studies will be necessary to determine the extent of this phenomenon across different species and conditions.

Moreover, changes to transcription[10] or RNA processing[11] can also change start-codon use. Are those mechanisms used in combination with uAUG–dsRNA cooperation? The authors report that three RH37-like RNA helicases are involved in unwinding dsRNA structures associated with uAUG start codons. This finding raises fresh questions. Are the functions of these helicases the same, or do they have different specificities? How are they regulated? Luckily, Xiang and colleagues showcase what approaches could be used to answer these questions.

**Yizhu Lin** and **Stephen N. Floor** are in the Department of Cell and Tissue Biology, University of California, San Francisco, San Francisco, California 94143, USA. S.N.F. is also at the Helen Diller Family Comprehensive Cancer Center, University of California, San Francisco.
e-mail: stephen@floorlab.org

1. Xiang, Y. *et al. Nature* **621**, 423–430 (2023).
2. Calvo, S. E., Pagliarini, D. J. & Mootha, V. K. *Proc. Natl Acad. Sci. USA* **106**, 7507–7512 (2009).
3. Dever, T. E., Ivanov, I. P. & Hinnebusch, A. G. *Genes Dev.* **37**, 474–489 (2023).
4. Hinnebusch, A. G. *Annu. Rev. Biochem.* **83**, 779–812 (2014).
5. Kozak, M. *Proc. Natl Acad. Sci. USA* **87**, 8301–8305 (1990).
6. Wang, J. *et al. Cell* **185**, 4474–4487 (2022).
7. Guenther, U.-P. *et al. Nature* **559**, 130–134 (2018).
8. Bohnsack, K. E., Yi, S., Venus, S., Jankowsky, E. & Bohnsack, M. T. *Nature Rev. Mol. Cell Biol.* https://doi.org/10.1038/s41580-023-00628-5 (2023).
9. Ryan, C. S. & Schröder, M. *Front. Cell Dev. Biol.* **10**, 1033684 (2022).
10. Cheng, Z. *et al. Cell* **172**, 910–923 (2018).
11. Hossain, M. A. *et al. Mol. Cell* **62**, 346–358 (2016).

Astrophysics

# Interstellar dust revealed by light from cosmic dawn

### Xuejuan Yang and Aigen Li

The obscuration of light from a distant galaxy has raised the possibility that a type of carbon dust existed in the earliest epochs of the Universe — challenging the idea that stars had not yet evolved enough to make such material. 

The space between stars is full of fine solid particles that range in size from several ångströms to a few micrometres. This 'interstellar dust' is a key component of galaxies[1] both near and far, and it dims the ultraviolet and optical light that arrives at Earth from these galaxies. The way that the dimming varies with wavelength is captured in 'extinction curves' that contain clues about the dust's composition and size. Curves for the Milky Way and other local galaxies exhibit a prominent bump at 2,175 Å, but some of these galaxies do not have this bump, and the same is often assumed of galaxies far away from Earth[2]. It therefore comes as a surprise that there is a pronounced extinction bump for a galaxy that lies in the distant reaches of the Universe, as Witstok *et al.*[3] report on page 267.

First discovered in the Milky Way nearly six decades ago[4], the 2,175 Å extinction bump is generally thought to be caused by nanoparticles containing aromatic carbon[5] — carbon atoms arranged in flat hexagonal rings. Graphite is the most stable form of pure aromatic carbon, and extraterrestrial graphite grains formed in the outflows of old, evolved stars and in the ejecta from supernovae have been found in primitive meteorites[6]. These remnants of the birth of the Solar System have remained essentially unchanged since their formation 4.6 billion years ago.

Another common type of aromatic material is a polycyclic aromatic hydrocarbon (PAH), which is composed of multiple hexagonal benzene rings that come together to form flat sheets edged with hydrogen atoms (Fig. 1). On Earth, PAHs are a common by-product of the incomplete combustion of organic matter, and are typically found in car exhaust and grilled meat. But extraterrestrial PAHs are abundant and widespread in the Universe — so much so that tell-tale signatures of PAHs are evident in light emitted at wavelengths of 3.3, 6.2, 7.7, 8.6, 11.3 and 12.7 μm (ref. 7).

Two other forms of carbon have also been detected in interstellar space — nanodiamonds and fullerenes. However, these materials do not produce an absorption band resembling the 2,175 Å bump, and neither is aromatic[8]. By contrast, aromatic materials (in particular, nanoscale graphite grains and PAHs) do exhibit such a band[5]. These materials are therefore widely considered to be responsible for the extinction bump observed in many galaxies.

Witstok and colleagues' detection of a 2,175 Å bump was made with the James Webb Space Telescope (JWST), as part of the JWST Advanced Deep Extragalactic Survey (JADES). The target galaxy, JADES-GS-z6-0, was observed by JWST as it existed 13 billion years ago; in other words, the light received by the telescope was radiated by JADES-GS-z6-0 just 800 million years after the Big Bang, during a period known as the cosmic dawn. The detection therefore indicates that aromatic nanoparticles were already pervasive in the Universe at the cosmic dawn — a remarkable finding with far-reaching astrophysical and astrochemical implications.

The origin of aromatic nanoparticles (and interstellar dust in general) in JADES-GS-z6-0 is puzzling. If star formation began when the Universe was about 400 million years old, Witstok and colleagues' detection comes from a time when even the oldest stars were only 400 million years old. At this point, low- to intermediate-mass stars (those about 0.5 to 8 times the mass of the Sun) would not have had enough time to evolve to the phase during which stars can produce dust, known as the asymptotic giant branch (AGB) phase.

In the Milky Way, AGB stars dominate the production of stardust and it typically takes them at least one billion years to evolve to this state. Because only stars with masses at least eight times that of the Sun would have been able to evolve so quickly, it is often suggested that supernovae were responsible for the dust in the first billion years of cosmic time.

These stellar explosions certainly make graphite — as shown by supernova-condensed graphite grains that have been identified in primitive meteorites[6] — but it is not clear whether PAHs form in supernova ejecta. The graphite grains found in meteorites are also too large (1–20 μm in diameter) to produce an absorption band around 2,175 Å (ref. 9). If the extinction bump seen in JADES-GS-z6-0 is indeed indicative of graphite nanoscale grains, and if these grains originated from supernovae, then there must be some mechanism through which micrometre-sized graphite grains can fragment into nanoparticles after being injected into the interstellar medium (Fig. 1).

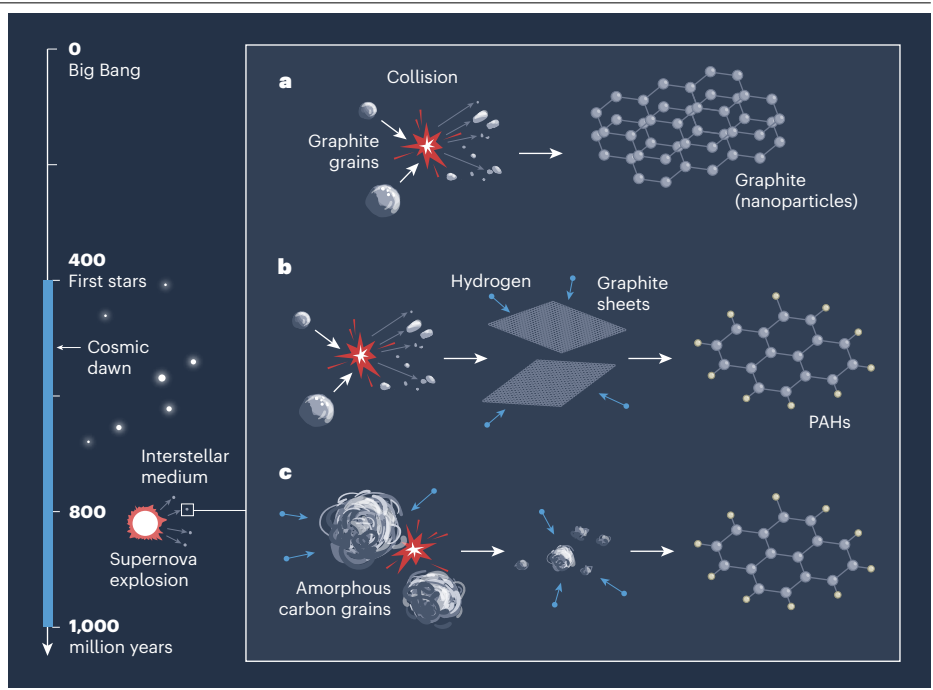

**Figure 1 | Carbon nanoparticles at the cosmic dawn.** The first stars formed at the start of the cosmic dawn, which began around 400 million years after the Big Bang and ended about 600 million years later. Witstok *et al.*[3] measured light emitted from the galaxy JADES-GS-z6-0 during this period and detected a feature at 2,175 Å that is generally attributed to aromatic carbon nanoparticles (containing flat hexagonal rings), such as graphite and polycyclic aromatic hydrocarbons (PAHs). At such an early epoch of cosmic time, these grains could not have been made in evolved stars; they must have originated in supernova ejecta or in the interstellar medium. **a**, Graphite grains formed in ejecta could have been blown into the interstellar medium and then shattered into nanoparticles through collision. **b**, Sheets of graphite generated from collisions could have transformed into PAHs after being bombarded by hydrogen in the interstellar medium. **c**, Amorphous carbon, formed in ejecta and injected into the interstellar medium, could have reacted with hydrogen and shattered, producing PAHs.

This fragmentation could occur if the grains shattered on collision. It has been argued that such destruction would have happened so fast that the interstellar dust could not have come from stellar outflows or supernova ejecta, and instead must have originated in the interstellar medium[1]. Supernovae could have provided 'seed' dust to initiate the growth of graphite grains in the interstellar medium, and this graphite could then have shattered, offering a viable mechanism for making graphite nanoparticles. However, given its layered structure, graphite could also have fragmented into planar graphene sheets, which, when bombarded by reactive atomic hydrogen in the interstellar medium, might then have transformed into PAHs (ref. 10).

Model simulations of dust condensation in primordial supernovae in the early Universe have found that amorphous carbon (a carbon material that lacks a well-defined structure) can also form in supernova ejecta[11]. When amorphous carbon is injected into the interstellar medium, it reacts with hydrogen atoms, and its hydrogenated form could also lead to the production of PAHs (through shattering) in the interstellar medium[12].

A study of the galaxy SPT0418-47 (observed less than 1.5 billion years after the Big Bang) showed that the spatial distribution of light from PAHs differs from that of larger dust particles, suggesting that the two particle types are not found in the same location[13]. But this conclusion seems inconsistent with the scenario that PAHs result from the fragmentation of graphite grains or amorphous carbon. For this reason, identifying the cause of the extinction bump detected in JADES-GS-z6-0 would provide key insights into the destruction and growth of dust in the early Universe.

The spectral profile of the extinction bump seen in JADES-GS-z6-0 is also interesting in its own right. The extinction bump for the Milky Way peaks at 2,175 Å for nearly all sources[14], but the JADES-GS-z6-0 bump occurs at a considerably longer wavelength, of about 2,263 Å, and it is substantially narrower than those of the Milky Way sources. In this sense, PAHs are more likely to be responsible for the bump than is graphite, because increased graphite grain size should shift the bump to longer wavelengths. However, the bump would then be broader than that observed by Witstok and colleagues. That said, mixtures of PAHs of different sizes and structures could produce a bump with a range of peak wavelengths and widths[5].

The metallicity of JADES-GS-z6-0 (the abundance of elements heavier than helium) is only 0.2–0.3 times that of the Sun[3], which would seem to contradict the general idea that PAHs are deficient in low-metallicity galaxies[7]. Indeed, the local galaxy known as the Small Magellanic Cloud has a low metallicity, similar to that of JADES-GS-z6-0, and essentially no extinction bump, as well as weak PAH emission[15]. The detection of an extinction bump in other low-metallicity galaxies would therefore provide crucial insight into how the bump is related to PAHs and to metallicity. Indeed, close positive correlations have been found between the bump, PAH emission light and stellar mass (which is indicative of metallicity) for 86 galaxies that were observed at the cosmic noon — a time about 3 billion years after the Big Bang, when star and galaxy formation were at their peak[16].

One way to examine whether PAHs are indeed present in JADES-GS-z6-0 is to search for their thermal-emission signatures at wavelengths of 3.3, 6.2, 7.7, 8.6, 11.3 and 12.7 μm (ref. 7). The expansion of the Universe stretches the wavelength of the light emitted by JADES-GS-z6-0, so the 3.3 μm band (which is stretched out to 25.5 μm) is the only one accessible to JWST. Detecting this band in JADES-GS-z6-0 would help scientists to determine the rate of star formation for young galaxies at a time when the Universe was in its infancy. Searching for signatures of PAHs in other galaxies will help astronomers to chart a timeline of cosmic star-formation history[17] — to which Witstok and colleagues have just made an important contribution.

**Xuejuan Yang** is in the School of Physics and Optoelectronics, Xiangtan University, 411105 Xiangtan, Hunan Province, China. **Aigen Li** is in the Department of Physics and Astronomy, University of Missouri, Columbia, Missouri 65211, USA.
e-mails: xjyang@xtu.edu.cn; lia@missouri.edu

1. Draine, B. T. in *Cosmic Dust — Near and Far* (eds Henning, T., Grün, E. & Steinacker, J.) **414**, 453–472 (Astron. Soc. Pacific Conf. Ser., 2009).
2. Salim, S. & Narayanan, D. *Annu. Rev. Astron. Astrophys.* **58**, 529–575 (2020).
3. Witstok, J. *et al. Nature* **621**, 267–270 (2023).
4. Stecher, T. P. *Astrophys. J.* **142**, 1683–1684 (1965).
5. Lin, Q., Yang, X. J. & Li, A. *Mon. Not. R. Astron. Soc.* **525**, 2380–2387 (2023).
6. Nittler, L. R. & Ciesla, F. *Annu. Rev. Astron. Astrophys.* **54**, 53–93 (2016).
7. Li, A. *Nature Astron.* **4**, 339–351 (2020).
8. Ma, X.-Y., Zhu, Y.-Y., Yan, Q.-B., You, J.-Y. & Su, G. *Mon. Not. R. Astron. Soc.* **497**, 2190–2200 (2020).
9. Draine, B. T. & Malhotra, S. *Astrophys. J.* **414**, 632–645 (1993).
10. Jones, A. P., Tielens, A. G. G. M. & Hollenbach, D. J. *Astrophys. J.* **469**, 740–764 (1996).
11. Todini, P. & Ferrara, A. *Mon. Not. R. Astron. Soc.* **325**, 726–736 (2001).
12. Scott, A., Duley, W. W. & Pinho, G. P. *Astrophys. J.* **489**, L193–L195 (1997).
13. Spilker, J. S. *et al. Nature* **618**, 708–711 (2023).
14. Wang, Q., Yang, X. J. & Li, A. *Mon. Not. R. Astron. Soc.* **525**, 983–993 (2023).
15. Sandstrom, K. M. *et al. Astrophys. J.* **715**, 701–723 (2010).
16. Shivaei, I. *et al. Mon. Not. R. Astron. Soc.* **514**, 1886–1894 (2022).
17. Xie, Y. & Ho, L. C. *Astrophys. J.* **884**, 136 (2019).

The authors declare no competing interests.

## Biotechnology

# A tool for optimizing messenger RNA sequence

### Anna K. Blakney

With messenger RNA therapeutics being developed for uses beyond vaccines, problems of RNA instability must be addressed. A new algorithm optimizes mRNA sequence for both stability and the encoding of amino acids. 

The SARS-CoV-2 pandemic greatly accelerated the development of messenger RNA technology by demonstrating the safety, effectiveness and scalability of mRNA-based vaccines[1]. On page 396, Zhang *et al.*[2] describe a tool that provides researchers with new ways to design mRNA sequences.

Messenger RNA is now used for various applications beyond vaccines for infectious diseases; these include treatments such as antibody therapy, cell therapies and personalized vaccines to boost immune responses to cancer. However, many challenges remain, including overcoming the low efficiency of mRNA-delivery systems and the need to modulate the immune response that the host generates as a result of this foreign RNA (immunogenicity).

RNA instability is the main hurdle to the development of mRNA medicines. For mRNA to be effective therapeutically, it must reach the cytoplasm of a target cell and persist there for long enough to express sufficient protein before being degraded. The sequence of the RNA affects its stability and the amount of protein expressed[3], but the specific features of RNA that govern these characteristics are poorly understood.

RNA stability must also be addressed in the context of the global distribution of vaccines. RNA is less stable at higher temperatures, necessitating its storage in freezers. Improving the stability of RNA might reduce its degradation during transport and storage.

One aspect that makes optimizing RNA sequences particularly difficult is the sheer number of potential sequence variations, which makes it impossible to test them all experimentally. For example, there are approximately $10^{632}$ possibilities in terms of RNA sequences that can encode the spike protein of SARS-CoV-2, which is the target of mRNA COVID-19 vaccines.

Zhang and colleagues introduce a powerful algorithm called LinearDesign that can

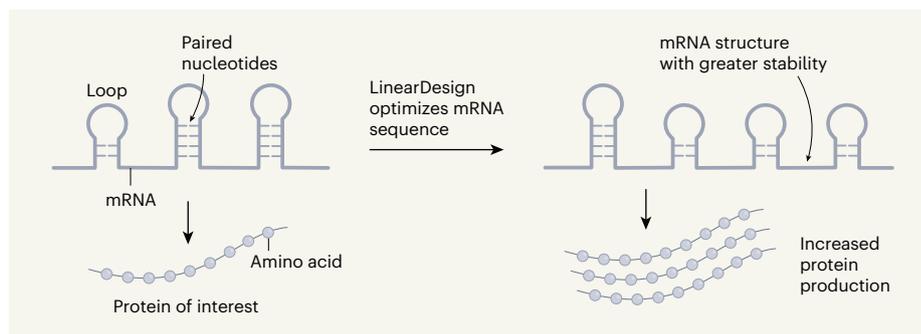

**Figure 1 | An algorithm that aids the design of messenger RNA sequences.** In most cases, various nucleotide sequences can encode a given amino acid. A protein can therefore be encoded by many possible mRNA sequences. These sequences can differ in their stability (which is affected by RNA structural elements such as loops that arise from nucleotide pairing) and the amount of protein that is made from the mRNA. Zhang *et al.*[2] present a tool for mRNA design called LinearDesign that can optimize mRNA selection. Having a stable sequence can help to limit mRNA loss through degradation.